\documentclass{iopjournal}

\usepackage{amsmath,amssymb}
\usepackage{bm}
\newcommand{\taup}{\tau_\omega}
\newcommand{\Eel}{E^{\rm elastic}}
\usepackage{cite}
\usepackage{xcolor}
\def\red{\color{black}}

\begin{document}

\articletype{Paper}

\title{Emergent odd viscoelasticity in chiral soft glassy materials}

\author{Debarghya Banerjee$^1$\orcid{0000-0003-3821-0874} and Peter Sollich$^2$\orcid{0000-0003-0169-7893}}

\affil{$^1$ Institut für Theoretische Physik, University of Göttingen, Friedrich-Hund-Platz 1, 37077 Göttingen, Germany}

\affil{$^1$ Ecole Polytechnique de l’Universit\'e de Nantes, LTeN, U6607, Nantes Universit\'e, Nantes, France}

\affil{$^2$ Institut für Theoretische Physik, University of Göttingen, Friedrich-Hund-Platz 1, 37077 Göttingen, Germany}

\affil{$^2$ Department of Mathematics, King’s College London, London WC2R 2LS, United Kingdom}

\affil{$^2$ Author to whom any correspondence should be addressed}

\email{peter.sollich@uni-goettingen.de}


\begin{abstract}
Rheological properties of chiral active materials have been an important area of research in the recent past, in particular regarding odd terms in their mechanical response. While much progress has been made in the study of odd viscous fluids and odd elastic solids, there is still a lack of understanding of odd {\em viscoelastic} responses. We introduce a chiral soft glassy rheology model to understand the emergence and nature of such odd viscoelastic responses in a class of amorphous solids. We use this model, which effectively considers an ensemble of actively rotating inclusions in a glassy matrix, to study the linear stress response to steady and oscillatory shear flows. For steady shear we find an odd viscosity that, non-trivially, grows as the active rotation frequency $\Omega$ decreases. In oscillatory shear we find an odd viscoelastic spectrum with a non-trivial dependence on the driving frequency $\omega$, combining resonance effects around $\omega= 2\Omega$ with glassy power laws at larger $\omega$.
\end{abstract}

\section{Introduction}

In simple fluids and elastic solids, the constitutive relations can be cast in the form of a linear equation connecting stress and strain-rate or strain, respectively~\cite{landau1986theory,landau2013fluid,chaikin1995principles}. However, this description does not explain the mechanical response of more realistic and complex materials~\cite{christensen2012theory,waigh2005microrheology,wu2007density,krishnan2010rheology,barnes1989introduction}, which exhibit viscoelastic behaviour. {\red A wide range of work on active biological matter~\cite{prost2015active,kruse2004asters,jawerth2020protein,ranft2010fluidization} has demonstrated such a viscoelastic response.} Amorphous systems, in particular, often show in their viscoelastic response a broad spectrum of relaxation times that is indicative of slow, glassy dynamics. Experiments have shown that such glassy behaviour arises in a wide range of biological systems including embryonic tissues~\cite{Schotz2013}, cell division and death~\cite{matoz2017cell} and chloroplasts in plants~\cite{Schramma2023}. Active glasses~\cite{janssen2019active,mandal2020multiple,szumera2005spectroscopic,yoshida2024structural,paul2023dynamical,ghosh2025elastoplastic,mandal2021shear} have been used to model many of these experimentally observed systems. As a consequence of their activity, such glassy systems can also exhibit chirality, which endows them with unique properties including e.g.\ fluidization by a ``hammering mechanism''~\cite{debets2023glassy}. Systems of active rotors, which are objects that generate mechanical rotation by burning e.g.\ chemical fuel, are an important example of such chiral active matter~\cite{marchetti2013hydrodynamics,ramaswamy2010mechanics,julicher2018hydrodynamic,bar2020self,fatemi2023optimal,liebchen2017collective}. The mechanical response functions of these materials have terms that are classified as ``odd'', and odd viscosity~\cite{avron1998odd,banerjee2017odd,banerjee2021active,fruchart2023odd,surowka2022odd,scheibner2020odd,banerjee2022hydrodynamic,souslov2019topological,duclut2024probe,Lier2023,Starzewski2024,ganeshan2017odd,samanta2022role,markovich2021odd,reichhardt2022active,yuan2023stokesian,markovich2024nonreciprocity} and odd elasticity~\cite{scheibner2020odd,fruchart2023odd,Starzewski2024,surowka2022odd,braverman2021topological,ishimoto2022self,fossati2024odd,kole2021layered,huang2023odd} have seen intense research activity  in recent years. Odd response functions are typically characterised by the breaking of the major symmetry of the fourth rank tensors, e.g.\ for the viscosity tensor $\eta_{ijkl}$, the odd component is defined as the one with the property $\eta^{\rm odd}_{ijkl} = -\eta^{\rm odd}_{klij}$. Experimentally, the presence of active rotation and the resulting odd viscosity have been shown e.g.\  in spinning colloidal magnets~\cite{soni2019odd}. More recent work on active rotation and its consequences for odd responses has studied systems with a range of structures and microscopic dynamics~\cite{hargus2024odd,markovich2024nonreciprocity,Matus2024}. In disordered materials, odd elastic responses arise in systems exhibiting Cosserat elasticity {\red -- an elastic theory of solids where the stress is coupled to an angle field, making it useful in describing structured materials and metamaterials --} in the presence of chiral active torques~\cite{Lee2025}. 

Our aim in this paper is to study odd {\em viscoelastic} responses in chiral active matter, in particular in the presence of structural disorder that generates glassy dynamics. For this purpose we propose a new chiral soft glassy rheology (SGR) model, building on the SGR approach for understanding viscoelastic properties and other rheological features in passive glassy materials~\cite{sollich1997rheology,sollich1998rheological,cates2004tensorial,bouchaud1992weak,bouchaud1995aging,sollich2006soft,bonn2002rheology,sollich2012thermodynamic,divoux2013rheological,fielding2014shear,voigtmann2014nonlinear}. In the original SGR model, a system is conceptually divided into mesoscopic elements and its dynamics described in terms of the distribution of local strains and yield energies. This distribution evolves in time by elastic loading due to external strain, and by yielding once local strains become of the order of the local yield strain. The combination of these two processes allows one to explain mechanical behaviour, such as the frequency dependence of elastic moduli, for a wide range of glassy materials. In the broader context of elastoplastic models~\cite{Nicolas2018}, the SGR model can be viewed as a mean-field approximation that neglects spatial correlations, and represents interactions by stress propagation between local elements as mechanical noise.

\begin{figure*}[htbp!]
    \centering
    \includegraphics[width=\linewidth]{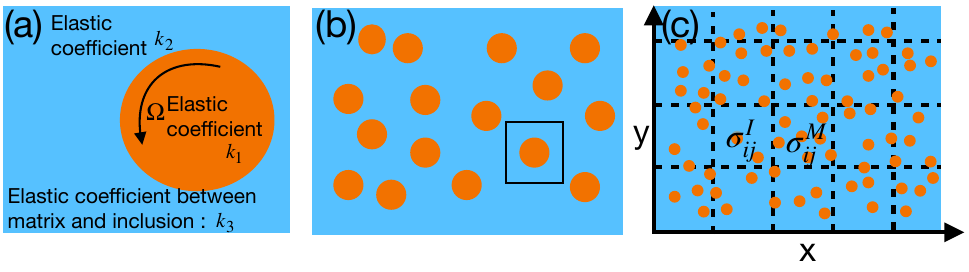} 
    \caption{{\bf Chiral SGR model} -- (a) Sketch of single mesoscopic element with an actively rotating inclusion (orange) inside a glassy matrix (blue). The corresponding elastic coefficients governing the shear response are indicated. (b) Sketch of the overall system of many inclusions distributed randomly in the matrix. (c) Pictorial representation of the spatial averaging over inclusions to obtain continuum equations for the stress of the inclusions ($\sigma^I_{ij}$) and of the matrix ($\sigma^M_{ij}$). 
    The original SGR model is recovered when the active rotation rate $\Omega$ is set to zero.
    }
    \label{fig:schematic}
\end{figure*}

To adapt the SGR model to chiral active matter, we again consider mesoscopic elements of the system, each containing a rotating ``inclusion'' surrounded by a glassy ``matrix''. Fig.~\ref{fig:schematic}(a) illustrates this mesoscopic element of the chiral SGR model, for the two-dimensional case that we focus on throughout. As in the original SGR model, we assume that the mesoscopic elements are large enough for local stresses and strains to be defined, yet small enough compared to the system size that we can eventually describe the macroscopic behaviour by suitable averaging. The key parameter for the chiral activity of the system will be the angular frequency $\Omega$ with which the inclusions rotate. We construct below the full master equation for the chiral SGR model, which requires us to track tensorial strains of both the inclusion and the matrix part of each mesoscopic element. A pre-averaging approximation that becomes exact for small deformations~\cite{fielding2020elastoviscoplastic} then allows us to derive {\red closed} equations for the inclusion and matrix stresses, {\red conditional on the local yield energy of the element. Finally, we coarse-grain in space across local regions containing many elements, in order to obtain a continuum description. In deriving this hydrodynamic theory we assume that the inclusions are distributed randomly, as is appropriate for a disordered, glassy material. (Additional terms could appear in the hydrodynamic theory if e.g.\ the inclusions had some spatial periodicity in their arrangement.)} We obtain the resulting stresses for steady state flows and find that these contain an odd viscosity term. This becomes larger as $\Omega$ decreases, a nontrivial effect resulting from the interplay of active chiral motion and glassy dynamics. Finally we consider oscillatory flows and derive the viscoelastic spectrum, which again contains an odd component with a frequency dependence exhibiting both (modified) glassy power laws and resonance-like effects when the driving frequency is of the order of $\Omega$.

\section{Matrix and inclusion elasticity}

As in the original SGR model we assign an elastic (free) energy to deformations of the matrix and inclusion parts of each mesoscopic element (Fig.~\ref{fig:schematic}(a)). Denoting the strain tensor of the inclusion by $l_{ij}^I$ ($i,j\in \{x,y\}$) and that of the matrix by $l_{ij}^M$, we assume linear elasticity for simplicity and so write a general quadratic form for the elastic energy,
\begin{align}
    E^{\rm elastic} = \frac{k_1}{2}\, 
    l_{ij}^I l_{ij}^I 
    + \frac{k_2}{2}\,
    l_{ij}^M l_{ij}^M 
    + k_3\, 
    l_{ij}^M l_{ij}^I 
\label{eq:energy}
\end{align}
where here and below repeated indices are summed over. The elastic constants $k_1$ and $k_2$ are taken to include the relevant factors from the respective areas of the inclusion and matrix part of the mesoscopic element, while the $k_3$-term arises physically from the energy cost of potential strain mismatch between inclusion and matrix, in the spirit of an Eshelby inclusion~\cite{eshelby1957determination,eshelby1959elastic,markenscoff1997shape}. Because of this mismatch the strain fields in the matrix part of an element will not generally be spatially homogeneous and $l_{ij}^M$ should accordingly be interpreted as the matrix strain {\red near} the boundary of the mesoscopic element. {\red We have assumed the material to the locally isotropic in writing Eq.~(\ref{eq:energy}). To simplify further} and as we will focus exclusively on shear deformations below, we have not separated out the shear and compression components of the strain tensors in writing $E^{\rm elastic}$; the elastic constants $k_1$, $k_2$, $k_3$ are those for shear deformations and therefore proportional to the relevant shear moduli. {\red Under these assumptions, Eq.~(\ref{eq:energy}) is the most general quadratic form for the combined elastic energy of the mesoscopic element.}
Using the above expression for the elastic energy we can compute the corresponding stress tensors: 
\begin{align}
    &  \frac{\partial E^{\rm elastic}}{\partial l_{ij}^{I}}  = k_1 l_{ij}^{I} + k_3 l_{ij}^{M} , \qquad \frac{\partial E^{\rm elastic}}{\partial l_{ij}^{M}} = k_2 l_{ij}^{M}  + k_3 l_{ij}^{I}\ .
\end{align}
We then assume that the total stress of the local material element, consisting of an inclusion and a local region of matrix around it, is given by\footnote{Somewhat more generally one could define the stress as $\sigma_{ij} = \xi\,{\partial E^{\rm elastic}}/{\partial l_{ij}^{I}} + {\partial E^{\rm elastic}}/{\partial l_{ij}^{M}}$, where the factor $\xi$ is related to the amount of slip on the boundary between the rotating inclusion and the matrix. We do not pursue this here as it only results in trivial modifications of our results by factors of $\xi$.}
\begin{align}
    \sigma_{ij} &= \left \langle \frac{\partial E^{\rm elastic}}{\partial l_{ij}^{I}} \right\rangle + \left \langle \frac{\partial E^{\rm elastic}}{\partial l_{ij}^{M}} \right \rangle = (1+k_3/k_1)\sigma_{ij}^I + (1+ k_3/k_2)\sigma_{ij}^M
\end{align}
where
\begin{equation}
\sigma_{ij}^I = k_1 \langle l_{ij}^{I} \rangle, \qquad 
    \sigma_{ij}^M = k_2 \langle l_{ij}^{M} \rangle\ .
\end{equation}
We assume finally that the strain tensors are related to the corresponding {\red elastic} deformation tensors, which we denote by $F_{ij}$, by the Green-Lagrange form\footnote{Other choices could be made but will give identical results in the regime of small deformations that we will concentrate on.} 
\begin{equation}
    l_{ij}^I = \frac{1}{2} (F_{ik}^I F_{jk}^I - \delta_{ij}), \qquad 
    l_{ij}^M = \frac{1}{2} (F_{ik}^M F_{jk}^M - \delta_{ij})\ .
    \label{Green-Lagrange}
\end{equation}
{\red Note that in the chiral SGR model described below, as in the original SGR model, deformations of local elements are taken to be reset when a plastic yield event occurs. This implies that the deformation tensors only carry information on elastic, rather than plastic, deformations.
Expressed within the Kr\"oner-Lee formalism~\cite{Lee1969,Kroner1960}, where the total deformation tensor $F^{T}_{ij} = F_{ik} F^{P}_{kj}$ is decomposed into an elastic deformation $F_{ij}$ and a plastic deformation $F^{P}_{ij}$, our deformation tensors thus track only the first, elastic contribution.} 

\section{Chiral SGR model}

{\red The chiral SGR model that we introduce below builds on the original SGR model. In the latter, the dynamics of mesoscopic elements follows simple rules: while no plastic event occurs, the deformation of each element changes with the 
externally imposed flow. Plastic yield events then occur stochastically, with rates corresponding to activation by mechanical noise and an energy barrier that is the difference between the yield energy $E$ and the elastic energy $E^{\rm elastic}$ of the element. Finally, after a yield event the element is taken to have relaxed its stress and is also assigned a new, randomly drawn yield energy. The distribution of yield energies represents the structural disorder in the material and thus encodes its glassy properties.}

To construct the chiral SGR model, we similarly consider the dynamics of the deformation of a single mesoscopic element in an externally imposed flow, described by a velocity field with components $v_i(x,y,t)$. Such an imposed flow would in practice be generated by shear stresses applied at the boundary of the system, which we do not represent explicitly. 
As in the original SGR model, we assume that the deformation changes of the matrix and inclusion part of the mesoscopic element follow the externally imposed velocity gradient $\partial_k v_i$. In addition, however, the 
inclusion rotates actively with angular frequency $\Omega$ in the chiral SGR model. To account for this, we need to work explicitly with the deformation tensors, which we denote by $F_{ij}$. Their dynamics is, by the above assumptions,
\begin{equation}
\frac{d}{dt}F_{ij}^I = (-\Omega \epsilon_{ik}  + \partial_k v_i )F_{kj}^I, \quad 
\frac{d}{dt}F_{ij}^M = (\partial_k v_i)F_{kj}^M
\end{equation} 
where $\epsilon_{ij}$ is the two-dimensional Levi-Civita symbol ($\epsilon_{12}=-\epsilon_{21}=1$, $\epsilon_{11}=\epsilon_{22}=0$). In addition to these elastic deformations, we assume -- again as in the original SGR model -- that plastic yield events can take place once the local elastic energy $E^{\rm elastic}$ gets sufficiently close to a local yield energy $E$. The rate for such yield events is taken to have an activated form, $\Gamma_0 e^{-(E- E^{\rm elastic})/\mathcal{X}}$~\cite{sollich1997rheology,sollich1998rheological,cates2004tensorial}, with attempt frequency $\Gamma_0$. The effective temperature $\mathcal{X}$ represents mechanical noise arising by stress propagation (see e.g. Refs.~\cite{Nicolas2018,picard2004elastic}) from yield events taking place elsewhere in the material. After a yield event, the local yield energy $E$ is assumed to be drawn randomly from a distribution $\rho(E)$, which plays the role of a density of states or prior distribution. The deformations $\bm{F}^I=\{F^I_{ij}\}$, $\bm{F}^M=\{F^M_{ij}\}$ of inclusion and matrix in the mesoscopic element are similarly taken to be reset in a yield event, to values drawn from residual deformation distributions 
$\rho^I\!(\bm{F}^I)$ and $\rho^M\!(\bm{F}^M)$, respectively. One choice that can be made for the distribution function of strains is $\rho^I\!(\bm{F}^I)=\prod_{ij}\delta(F_{ij}^I-\delta_{ij})$ and similarly for $\rho^M\!(\bm{F}^M)$. This corresponds to full relaxation at yielding, as assumed in most of the existing SGR literature.

Putting everything together, the statistics $P(E,\bm{F}^I,\bm{F}^M,t)$ of mesoscopic elements, in a local region where the velocity gradient $\partial_k v_i$ can be taken as approximately spatially constant, evolves according to the master equation
\begin{align}
    \frac{\partial}{\partial t} P = 
    & - \frac{\partial }{\partial F_{ij}^I} \left( P (-\Omega \epsilon_{ik}  + \partial_k v_i )F_{kj}^I \right) -  \frac{\partial }{\partial F_{ij}^M}  (P (\partial_k v_i) F_{kj}^M) \nonumber \\
    & - \Gamma_0 e^{-(E- E^{\rm elastic} )/ \mathcal{X}}  P + \Gamma(t) \rho(E)
    \rho^I\!(\bm{F}^I)\rho^M\!(\bm{F}^M) 
    \ .
\label{eq:sgr}
\end{align}
The last term, which represents the statistics of mesoscopic elements after a yield, contains as prefactor the overall local yield rate $\Gamma(t) = \Gamma_0 \langle e^{-(E- E^{\rm elastic})/\mathcal{X}} \rangle_P$, thus ensuring probability conservation. Eq.~(\ref{eq:sgr}) can usefully be rewritten in a form that isolates the strain dependence of the yielding processes, as
\begin{align}
     \frac{\partial}{\partial t}{P} & + \frac{\partial}{\partial F_{ij}^I} \left( \left(-\Omega \epsilon_{ik} + \partial_k v_i \right) F_{kj}^I P  \right) + \frac{\partial}{\partial F_{ij}^M} ((\partial_k v_i) F_{kj}^M P ) \nonumber \\
    &= -\frac{1}{\tilde{\tau}} f(\{l_{ij}^I\},\{l_{ij}^M\}) P + \Gamma(t) \rho(E) \rho^I\!(\bm{F}^I)\rho^M\!(\bm{F}^M)
    \label{Chiral_SGR_1}
\end{align}
with the definitions
\begin{align}
    \tilde{\tau} = \tau_0 \exp \left( \frac{E}{\mathcal{X}}\right), \qquad
    f(\{l_{ij}^I\},\{l_{ij}^M\}) = \exp \left( \frac{E^{\rm elastic}}{\mathcal{X}}\right)
\end{align}
and $\tau_0=1/\Gamma_0$.

In analysing the chiral SGR model as defined by Eq.~(\ref{eq:sgr}), we will initially keep the density of states $\rho(E)$ arbitrary but eventually take it to be of the exponential form used in the original SGR model, {\red i.e.\ $\rho(E)=e^{-E}$ for $E>0$. In general the solution of Eq.~\eqref{full} of course also requires specification of an initial distribution $P(E, {\bm F}^I, {\bm F}^M,0)$ at time $t=0$. This would be important for studying e.g.\ aging effects at low $\mathcal{X}$, but in this paper we will only consider steady state properties for which the initial preparation of the system becomes irrelevant.}

For the residual deformation distributions we will assume that the average residual strain tensor after a yield vanishes,
$\int d {\bm F}^I\,l_{ij}^I\,\rho^I\!({\bm F}^I) = 0$, and similarly for 
$\rho^M\!({\bm F}^M)$.
 This property holds if, for example, the residual deformations are isotropically distributed -- making the average of $l_{ij}$ proportional to $\delta_{ij}$
-- and area-preserving so that the trace of the average has to vanish. It should be noted that in the chiral SGR model of Eq.~(\ref{eq:sgr}), all effects of chiral activity have been collected into the single parameter $\Omega$ characterising the rotation of the inclusions. This rotation could alternatively be considered as arising from an active torque, leading in an overdamped setting to the torque-balance equation $\tau^{\rm active} = \gamma \Omega$ and hence $\Omega = \tau^{\rm active}/ \gamma$~\cite{fily2012cooperative,leoni2011dynamics,van2016spatiotemporal}. The energy balance equation of the chiral SGR model can also be derived and turns out to have (for small strains, see Appendix~\ref{app:energy}) the same form as in conventional SGR.

\section{Pre-averaging approximation}

From Eq.~(\ref{eq:sgr}) we now want to derive hydrodynamic equations for the local stress, which is a linear combination of the inclusion and matrix stresses $\sigma^I_{ij}=k_1\langle l_{ij}^I\rangle$ and $\sigma^M_{ij}=k_2\langle l_{ij}^M\rangle$, respectively. In Fig.~\ref{fig:schematic}(b) we show a sketch of the overall system, where after local averaging (over regions containing many mesoscopic elements) one obtains continuum stress fields as indicated in Fig.~\ref{fig:schematic}(c). The relevant local averages are over $P(E,\bm{F}^I,\bm{F}^M,t)$ and in principle depend on all details of this distribution. To close the equations, we use a pre-averaging approach~\cite{fielding2020elastoviscoplastic} that replaces the strains $l^I_{ij}$ and $l^M_{ij}$ appearing in $E^{\rm elastic}$ by their averages. Given the quadratic strain dependence of $E^{\rm elastic}$, this approximation is exact to linear order in the strains. 

To construct this approximation, we separate the full probability $P(\bm{F}^I,\bm{F}^M,E,t)$ into 
the marginal distribution over yield energies, $G(E,t) = \int d\bm{F}^I\,d\bm{F}^M P(\bm{F}^I,\bm{F}^M,E,t)$, and the remaining conditional probability distribution $P_1(\bm{F}^I,\bm{F}^M|E,t)=P(\bm{F}^I,\bm{F}^M,E,t)/G(E,t)$ of deformations of local elements with a given yield energy $E$. Accordingly, we have $P = G(E,t) P_1$ and the master equation Eq.~(\ref{Chiral_SGR_1}) takes the form
\begin{align}
    & P_1\frac{\partial}{\partial t}G(E,t) + G(E,t) \frac{\partial}{\partial t}{P}_1 
    + G(E,t)  \frac{\partial}{\partial F_{ij}^I} \left( (\partial_k v_i) F_{kj}^I P_1 \right) -  G(E,t) \frac{\partial}{\partial F_{ij}^I} \left(\Omega \epsilon_{ik} F_{kj}^I  P_1 \right)
\nonumber \\
    & + G(E,t)  \frac{\partial}{\partial F_{ij}^M} \left((\partial_k v_i) F_{kj}^M  P_1 \right) = -\frac{G(E,t)}{\tilde{\tau}} f(\{l_{ij}^I\},\{l_{ij}^M\}) P_1 
    + \Gamma(t) \rho(E) 
    \rho^I\!(\bm{F}^I)\rho^M\!(\bm{F}^M)
    \label{full}
\end{align}
We now integrate over the local deformation variables $\bm{F}^I$ and $\bm{F}^M$ (see Appendix~\ref{app:G}) to obtain the time evolution of the distribution of yield energies:
\begin{align}
    \frac{\partial}{\partial t} G(E,t) = -\frac{G(E,t)}{\tilde{\tau}} \langle f(\{l_{ij}^I\},\{l_{ij}^M\}) \rangle_{E,t} + \Gamma(t) \rho(E)
    \label{G_dot}
\end{align}
where the average on the r.h.s.\ is over the conditional distribution $P_1(\bm{F}^I,\bm{F}^M|E,t)$, as indicated by the conditioning variables in the subscript. As expected, terms in Eq.~(\ref{full}) corresponding to local deformation changes do not contribute here, with $G(E,t)$ evolving only due to plastic yielding. 

In a similar fashion we can multiply Eq.~(\ref{full}) by $l_{ij}^I$ or $l_{ij}^M$ and then integrate over $\bm{F}^I$ and $\bm{F}^M$, to get equations of motion for the averages of the local strain tensors. We recall that the latter are related to the deformation gradient tensors by Eq.~{\eqref{Green-Lagrange}}. We begin with the somewhat simpler $l^M_{ij}$. Multiplying by this and integrating at fixed $(E,t)$, the first term on the l.h.s.\ of Eq.~\eqref{full} becomes $(\partial G/\partial t)\langle l^M_{ij}\rangle_{E,t}$ and the second one $G(\partial/\partial t)\langle l^M_{ij}\rangle_{E,t}$. Using the decomposition $P=GP_1$ introduced above, the average matrix stress can be written as
\begin{equation}
    \sigma^M_{ij}=k_2\langle l^M_{ij}\rangle = \int dE\,
    \sigma^M_{ij}(E,t)
\end{equation}
where
\begin{equation}
    \sigma^M_{ij}(E,t)=
    k_2 G(E,t)\langle l^M_{ij}\rangle_{E,t} 
\end{equation}
is the contribution to the stress from elements with yield energy $E$. The first two terms on the l.h.s.\ of Eq.~\eqref{full} that we have just worked out therefore combine to $(\partial/\partial t)\sigma^M_{ij}(E,t)/k_2$. 
The third and fourth terms {\red only give boundary terms at infinity and therefore vanish, using the same argument as in Appendix~\ref{app:G}}. On the r.h.s., the first term gives $-G\tilde{\tau}^{-1}\langle l_{ij}^Mf\rangle_{E,t}$ and the second one $\Gamma(t)\langle l^M_{ij}\rangle_\rho$ where the subscript $\rho$ indicates an average over the post-yield distribution.
But the average $\langle l^M_{ij}\rangle_\rho$ actually vanishes, provided the residual strains after yield are volume-preserving on average and randomly oriented as assumed above. {\red The calculation for the nontrivial fifth term on the l.h.s.\ of Eq.~\eqref{full} is shown in Appendix~\ref{app:premultiply}}. 
Combining then all terms and multiplying by $k_2$ gives the equation of motion for $\sigma^M_{ij}(E,t)$ as
\begin{equation}
(\partial/\partial t)\sigma^M_{ij}-\left[(\partial_k v_i) \sigma_{kj}^M+  (\partial_k v_j) \sigma_{ki}^M + k_2 G v_{ij}  \right]
= - k_2 G \tilde{\tau}^{-1}\langle l_{ij}^M f\rangle_{E,t}
\label{sigmaM_dot}
\end{equation}
where 
$v_{ij}=(\partial_i v_j + \partial_j v_i)/2$ is the symmetrized velocity-gradient tensor and we have dropped all arguments ($E,t$) for brevity.
A fully analogous calculation gives for the contribution to the inclusion stress of local elements with yield energy $E$, defined as
\begin{equation}
    \sigma^I_{ij}(E,t)=
    k_1 G(E,t)\langle l^I_{ij}\rangle_{E,t} \ ,
\end{equation}
the equation of motion
\begin{equation}
(\partial/\partial t)\sigma^I_{ij}-\left[ (\partial_k v_i) \sigma_{kj}^I + (\partial_k v_j) \sigma_{ki}^I + k_1 G v_{ij}  \right]
= - k_1 G \tilde{\tau}^{-1}\langle l_{ij}^I f\rangle_{E,t}
- \Omega (\epsilon_{ik} \sigma_{kj}^I  + \epsilon_{jk} \sigma_{ik}^I)\ .
\label{sigmaI_dot}
\end{equation}

So far everything is an exact consequence of the master equation Eq.~(\ref{full}), but the equations for $G(E,t)$, $\sigma^I_{ij}(E,t)$, $\sigma^M_{ij}(E,t)$ are not closed because of the averages over $P_1$ appearing on the r.h.s.\ of Eqs.~(\ref{G_dot},\ref{sigmaM_dot},\ref{sigmaI_dot}). To close the equations, we make the pre-averaging approximation from Ref.~\cite{fielding2020elastoviscoplastic}, in which $f(\{l_{ij}^I\},\{l_{ij}^M\})$ is replaced by $f(\{\langle l_{ij}^I\rangle\},\{\langle l_{ij}^M\rangle \})$ everywhere, the averages being conditional on $(E,t)$. In this way $f$ becomes just a function of $\{\sigma^I_{ij}\}$, $\{\sigma^M_{ij}\}$ and can be pulled out of all the averages so that e.g.\ the r.h.s.\ of Eq.~(\ref{sigmaM_dot}) becomes $-G\sigma^M_{ij}f\tilde\tau^{-1}=-G\sigma^M_{ij}/\tau$, with $\tau$ evaluated in the same pre-averaging approximation as $f$. 
In this way we obtain the following set of closed equations of motion:
\begin{eqnarray}
(\partial/\partial t) G &=& -\frac{G}{\tau} + \Gamma(t)\rho(E)
\label{G_SM}\\
     (\partial/\partial t) \sigma_{ij}^I - ({\red \partial_k v_i}) \sigma_{kj}^I - (\partial_k v_j) \sigma_{ki}^I &=&  k_1 G v_{ij} - \frac{\sigma_{ij}^I}{\tau} - \Omega (\epsilon_{ik} \sigma_{kj}^I  + \epsilon_{jk} \sigma_{ik}^I) \label{sigma1_SM} \\
     (\partial/\partial t) \sigma_{ij}^M - ({\red \partial_k v_i}) \sigma_{kj}^M - (\partial_k v_j) \sigma_{ki}^M &=& k_2 G v_{ij} - \frac{\sigma_{ij}^M}{\tau}\ .
     \label{sigma2_SM}
\end{eqnarray}
{\red We note that while the pre-averaging approximation is uncontrolled in general, it was shown in Ref.~\cite{fielding2020elastoviscoplastic} that for the original SGR model it is in fact exact for the linear response regime, which is the one we focus on in this paper.}

Our considerations so far have related to a region of the material at a fixed position in space; in a flow we need to replace $(\partial /\partial t)$ in the equations above by the advective derivative $D_t \equiv (\partial/\partial t) + ({\bm v} \cdot \nabla )$, giving
\begin{eqnarray}
D_t \sigma_{ij}^I -({\red \partial_k v_i}) \sigma_{kj}^I - (\partial_k v_j) \sigma_{ki}^I &=&  k_1 G v_{ij} - \frac{\sigma_{ij}^I}{\tau} - \Omega (\epsilon_{ik} \sigma_{kj}^I  + \epsilon_{jk} \sigma_{ik}^I)
\label{eq:continuumSGR1}\\
D_t \sigma_{ij}^M - ({\red \partial_k v_i}) \sigma_{kj}^M - (\partial_k v_j) \sigma_{ki}^M &=& k_2 G v_{ij} - \frac{\sigma_{ij}^M}{\tau}.
\label{eq:continuumSGR}
\end{eqnarray} 
These equations can be read as a modified form of the Maxwell model for viscoelasticity, with the last term in Eq.~\eqref{eq:continuumSGR1} accounting for the effect of the active rotation of the inclusion. {\red Note, however, that} in Eqs.~(\ref{eq:continuumSGR1},\ref{eq:continuumSGR}) all dynamical variables $\sigma^I_{ij}$, $\sigma^M_{ij}$, $G$ are {\red still} conditional on the yield energy $E$: total stresses are obtained in the end by integrating over $E$, {\red leading to nontrivial viscoelastic spectra when compared to those of a generic odd Maxwell model~\cite{banerjee2021active} (see Appendix~\ref{app:oddspectra}).} 
The quantities for different yield energies $E$ are coupled via the overall (local) yield rate
\begin{equation}
\Gamma(t) = \Gamma_0 \int dE~\,G(E,t)\langle f\rangle_{E,t}\tilde{\tau}^{-1} \to  
\Gamma_0 \int dE\,\frac{G(E,t)}{\tau}
\end{equation}
where pre-averaging has been used again in the last step.

\section{Steady shear flow}

We first consider the stress response to steady state, spatially uniform flows $v_i$, within a linear theory where convective and co-rotational terms can be discarded so that Eqs.~(\ref{G_SM},\ref{sigma1_SM},\ref{sigma2_SM}) simplify to
\begin{eqnarray}
(\partial/\partial t)G&=&-\frac{G}{\tau}+\Gamma(t)\rho(E)
\label{G_SM_lin}\\
     (\partial/\partial t)\sigma_{ij}^I &=&  k_1 G v_{ij} - \frac{\sigma_{ij}^I}{\tau} - \Omega (\epsilon_{ik} \sigma_{kj}^I  + \epsilon_{jk} \sigma_{ik}^I) \label{sigma1_SM_lin} \\
     (\partial /\partial t) \sigma_{ij}^M &=& k_2 G v_{ij} - \frac{\sigma_{ij}^M}{\tau}\ 
.\label{sigma2_SM_lin}
\end{eqnarray}
We recall once more that $G$, $\{\sigma^I_{ij}\}$ and $\{\sigma^M_{ij}\}$ all depend on $E$, while for a steady state situation they are of course independent of $t$. In the linear regime, given that $E^{\rm elastic}$ is quadratic in strains, we have $f=1$ to linear order so that $\tau=\tilde{\tau}=\tau_0 e^{E/\mathcal{X}}$. For brevity we therefore drop the tilde on $\tilde{\tau}$ below.

In a steady state the right hand sides of Eqs.~(\ref{G_SM_lin},\ref{sigma1_SM_lin},\ref{sigma2_SM_lin}) need to vanish, giving for $G$
\begin{equation}
G(E)= \Gamma \tau(E) \rho(E), \qquad \Gamma^{-1} = \int dE~\,{\tau(E)\rho(E)}\ .
\label{G_SM_ss}
\end{equation}
Here the steady state yield rate $\Gamma$ follows from the normalization of $G(E)$, which we recall is the distribution of yield energies across local elements. The results in Eq.~(\ref{G_SM_ss}) are, as expected, identical to those for the original SGR model~\cite{sollich1997rheology,sollich1998rheological}. 

For the steady state stresses, which we denote by $\Sigma$, one similarly finds the conditions 
\begin{eqnarray}
    \Sigma_{ij}^I &=& \tau \left[ k_1 G v_{ij} - \Omega (\epsilon_{ik} \Sigma_{kj}^I  + \epsilon_{jk} \Sigma_{ik}^I)\right], \label{sigmaI_steady}\\
    \Sigma_{ij}^M &=& \tau k_2 G v_{ij}.
\end{eqnarray}
Solving these linear equations explicitly yields for the inclusion stress 
\begin{eqnarray}
    \begin{pmatrix}
    \Sigma_{xx}^I \\
    \Sigma_{yy}^I \\
    \Sigma_{xy}^I 
    \end{pmatrix} 
    =\Sigma_0  \begin{pmatrix} 
    1 + 2\Omega^2 \tau^2 & 2 \Omega^2 \tau^2 & 2 \Omega \tau \\
    2\Omega^2 \tau^2 & 1 + 2 \Omega^2 \tau^2 & -2 \Omega \tau \\
    - \Omega \tau & \Omega \tau & 1        
    \end{pmatrix} \begin{pmatrix}
        v_{xx} \\
        v_{yy} \\
        v_{xy}
    \end{pmatrix}
\label{eq:steady_state}
\end{eqnarray}
with
\begin{equation}    
\Sigma_0 = \frac{k_1 G \tau}{1 + 4 \Omega^2 \tau^2}\ ,
\end{equation}
while for the matrix
\begin{equation}
    \Sigma_{ij}^M = k_2 G \tau v_{ij}\ . 
\end{equation}
The matrix stresses thus behave as in the original SGR model, with elements with higher yield energy $E$ carrying larger stresses because they are slower to yield (large $\tau$). The effects of the active rotation appear in the inclusion stress $\Sigma_{ij}^I$, with the nonzero entries in the third row and column of the matrix in Eq.~(\ref{eq:steady_state}) indicating, in addition to the usual shear viscosity, an odd viscous response. The dependence of the total odd viscosity on $\Omega$, obtained by integrating over $E$, will be analysed below.

One can ask about the stability of the steady state analysed above: small perturbations $\tilde\sigma^{I,M}_{ij}=\sigma^{I,M}_{ij}-\Sigma^{I,M}_{i,j}$ around it evolve as
\begin{align}
    & (\partial/\partial t) \tilde{\sigma}_{ij}^I = -\frac{1}{\tau} \tilde{\sigma}_{ij}^I - \Omega  (\epsilon_{ik} \tilde{\sigma}_{kj}^I  + \epsilon_{jk} \tilde{\sigma}_{ik}^I), \label{sigma1_linearized} \\
    & (\partial /\partial t) \tilde{\sigma}_{ij}^M = -\frac{1}{\tau} \tilde{\sigma}_{ij}^M\ .
\end{align}
This shows that $\sigma_{ij}^M$ is always stable and relaxes back to the steady state at rate $1/\tau$. For $\sigma_{ij}^I$, one finds by writing Eq.~\eqref{sigma1_linearized} as a linear equation for the vector $(\tilde{\sigma}^I_{xx},\tilde{\sigma}^I_{yy},\tilde{\sigma}^I_{xy})$ that the three eigenvalues are $-1/\tau$ and $-1/\tau\pm 2i\Omega$, giving two additional damped oscillatory relaxation modes arising from the active rotation frequency $\Omega$.

\section{Oscillatory shear flow}
\label{sec:oscillatory}

We next consider more generally an oscillatory flow with shear rate $v_{ij} = v^0_{ij}e^{i \omega t}$. The linear response of the matrix stress is then governed by $\partial_t \sigma_{ij}^M = k_2 G v^0_{ij}e^{i\omega t} - \sigma_{ij}^M/\tau$, leading to a Maxwell-type behaviour. The non-trivial inclusion stress obeys instead
\begin{align}
    & \partial_t \sigma_{ij}^I - k_1 G v^0_{ij} e^{i \omega t} = -\frac{\sigma_{ij}^I}{\tau} - \Omega (\epsilon_{ik} \sigma_{kj}^I + \epsilon_{jk} \sigma_{ik}^I)
    \label{ddt_sigmaI}
\end{align}
so that the three independent stress components $\sigma^I_{xx}$, $\sigma^I_{yy}$, and $\sigma^I_{xy}$ are coupled. After the decay of initial transients the solution is of the form $\sigma^I_{ij} = \sigma^{I,0}_{ij}
e^{i\omega t}$ with $\sigma^{I,0}_{ij}=
\eta_{ijkl} v^0_{kl}$
and $\eta_{ijkl}$ a frequency-dependent fourth-rank viscosity tensor. Dropping the 0-superscripts for brevity leads to the equations
\begin{align}
     & i\omega \sigma_{ij}^I =  k_1 G v_{ij} - \frac{\sigma_{ij}^I}{\tau} - \Omega (\epsilon_{ik} \sigma_{kj}^I  + \epsilon_{jk} \sigma_{ik}^I), \\
     & i\omega \sigma_{ij}^M = k_2 G v_{ij} - \frac{\sigma_{ij}^M}{\tau}\ .
\end{align}
Bringing the terms on the l.h.s.\ to the right, one finds the same equations as in steady shear (see previous section), except for the replacement of $1/\tau$ by $1/\taup=1/\tau+i\omega$; explicitly, one has $\taup = \tau/(1+i\omega\tau)$. Translating the steady state results to the current setting then gives
\begin{align}
    \begin{pmatrix}
    \sigma_{xx}^I \\
    \sigma_{yy}^I \\
    \sigma_{xy}^I 
    \end{pmatrix} =\frac{k_1 G \taup}{1 + 4 \Omega^2 \taup^2} \begin{pmatrix} 
    1 + 2\Omega^2 \taup^2 & 2 \Omega^2 \taup^2 & -2 \Omega \taup \\
    2\Omega^2 \taup^2 & 1 + 2 \Omega^2 \taup^2 & 2 \Omega \taup \\
     \Omega \taup & -\Omega \taup & 1        
    \end{pmatrix} \begin{pmatrix}
        v_{xx} \\
        v_{yy} \\
        v_{xy}
    \end{pmatrix}
    \label{eq:vis_m_response}
\end{align}
and 
\begin{align}
    \begin{pmatrix}
    \sigma_{xx}^M \\
    \sigma_{yy}^M \\
    \sigma_{xy}^M 
    \end{pmatrix} =k_2 G \taup \begin{pmatrix}
        v_{xx} \\
        v_{yy} \\
        v_{xy}
    \end{pmatrix}
\label{eq:vis_response}
\end{align}
From the last expressions one sees that, at the level of linear response, the stresses in the matrix exhibit a simple Maxwell behaviour (for, we recall, fixed $E$), with a complex viscosity proportional to $\taup =\tau/(1+i\omega\tau)$. 

For the inclusion stresses, the viscosity tensor in Eq.~(\ref{eq:vis_m_response}) has a more complicated form. However, this can be decomposed using partial fractions and written in terms of three frequency dependent viscosity terms $\eta_0$, $\eta_+$ and $\eta_-$ defined as
\begin{align}
    & \eta_0 = \frac{k_1 G \tau}{1 + i \omega \tau},  \qquad  
    \eta_\pm = \frac{k_1 G \tau}{2(1 + i \omega \tau  \pm 2 i \Omega \tau)}
    \label{eta0pm_def}
\end{align}
and corresponding to the three eigenvalues determining the stability of the unsheared steady state.
Rewriting Eq.~(\ref{eq:vis_m_response}) in terms of these functions we obtain 
\newcommand{\hf}{\frac{1}{2}}
\begin{align}
    \begin{pmatrix}
    \sigma_{xx}^I \\
    \sigma_{yy}^I \\
    \sigma_{xy}^I 
    \end{pmatrix} =\begin{pmatrix} 
    \hf(\eta_0 + \eta_+ + \eta_-) & \hf(\eta_0 - \eta_+ - \eta_-) & -i (\eta_+ -\eta_-) \\
    \hf(\eta_0- \eta_+ - \eta_-)& \hf(\eta_0 + \eta_+ + \eta_-) & i (\eta_+ -\eta_-) \\
     \frac{i}{2} (\eta_+ -\eta_-) & -\frac{i}{2} (\eta_+ -\eta_-) & \eta_+ + \eta_- 
    \end{pmatrix} \begin{pmatrix}
        v_{xx} \\
        v_{yy} \\
        v_{xy}
    \end{pmatrix}
    \label{eq:vis_m_response_decomposed}
\end{align}
From this we can now extract the frequency-dependent even shear
and odd viscosities~\cite{avron1998odd}, which we define as 
\begin{equation}
\eta_{\text{shear}} = \eta_{xyxy}, \qquad 
\eta_{\text{odd}} = \eta_{xyxx}
\end{equation}
where $\eta_{xyxy}$ etc are the relevant entries of the matrix in Eq.~(\ref{eq:vis_m_response_decomposed}). Explicitly we find 
\begin{align}
    & \eta_{\text{shear}} = \eta_+ + \eta_-
    \label{shear_via_plus_minus}
    \\
    & \eta_{\text{odd}} = \frac{i}{2} (\eta_+ - \eta_-)\ .
    \label{odd_via_plus_minus}
\end{align}
We note as an aside that one could also extract from the above calculation the bulk viscosity $\eta_{\text{bulk}} = \eta_{xxxx}= \eta_{yyyy}$, which in our analysis of incompressible flows plays no role. This bulk viscosity comes out here as $\eta_{\text{bulk}} = \eta_0 + \eta_+ + \eta_-$, but a consistent determination of the prefactors would require introducing within $\Eel$ separate elastic constants for shear and compression, respectively.  

So far our calculation applies to local elements with a given yield energy $E$, whose value enters through $G(E)$ and $\tau(E)=\tau_0\exp(E/\mathcal{X})$. Using the standard SGR assumption that the prior yield energy distribution is exponential, $\rho(E)=e^{-E}$ for $E>0$, one has from (\ref{G_SM_ss}) that 
$G(E)=(1-1/\mathcal{X})e^{-E(1-1/\mathcal{X})}$. Integrating $\eta_0$ from Eq.~(\ref{eta0pm_def}) over all $E$ gives then
\begin{equation}
    \eta_0^*(\omega) = \int_0^\infty dE\,\eta_0(E,\omega) = k_1 \int_0^\infty dE\, \frac{G(E)\tau(E)}{1+i\omega \tau(E)} = k_1 \int_{\tau_0}^\infty d\tau\, P_{\rm ss}(\tau) \frac{\tau}{1+i\omega \tau} 
    \label{eta0_int}
\end{equation}
Here we have explicitly written the arguments $E$ and $\omega$ omitted above, and added an asterisk as superscript on the integrated version to conform with standard notation for complex viscosities. In the last step we have also changed integration variable from $E$ to $\tau$ and identified the steady state distribution of $\tau$,
\begin{equation}
    P_{\rm ss}(\tau) = (\mathcal{X}-1)\tau_0^{\mathcal{X}-1}\tau^{-\mathcal{X}}\ .
    \label{Pss_tau}
\end{equation}
One observes that the result Eq.~(\ref{eta0_int}) is exactly the frequency-dependent viscosity of the standard SGR model~\cite{sollich1997rheology,sollich1998rheological}, except that in the latter the modulus prefactor and the timescale $\tau_0$ are normally set to unity and the result is expressed in terms of the corresponding viscoelastic modulus $G^*(\omega)=i\omega \eta^*(\omega)$. The latter is seen from Eq.~(\ref{eta0_int},\ref{Pss_tau}) to be a superposition of Maxwell modes with different relaxation times $\tau$, which are distributed according to the power law spectrum $P_{\rm ss}(\tau)$.  

For the viscosities $\eta_\pm$ from Eq.~(\ref{eta0pm_def}) involving the active rotation velocity $\Omega$, one finds by an exactly analogous calculation
\begin{equation}
    \eta_\pm^*(\omega) = k_1 \int_{\tau_0}^\infty d\tau\, P_{\rm ss}(\tau) \frac{\tau}{2(1+i\omega \tau\pm 2i\Omega\tau)} = \frac{1}{2}\eta_0^*(\omega\pm 2\Omega) 
    \label{etapm_int}
\end{equation}
We can finally use the above expressions to write down the overall shear and odd viscosities as a function of frequency, in the form
\begin{eqnarray}
\eta^*_{\text{shear}}(\omega) &=& 
    \frac{1}{2} [\eta^*_0(\omega+2\Omega) + \eta^*_0(\omega-2\Omega)]
    \label{eta_shear}\\
\eta^*_{\text{odd}}(\omega) &=& \frac{i}{4} [\eta^*_0(\omega+2\Omega) - \eta^*_0(\omega-2\Omega)] \ .
\label{eta_odd}
\end{eqnarray}
So far we have evaluated the contribution to the dynamical viscosity from the inclusion. From Eq.~(\ref{eq:vis_response}) one reads off that the role of the matrix is rather simpler: it makes no contribution to the odd viscosity, while to the shear viscosity it adds a term proportional to $\eta_0^*(\omega)$ (see Appendix~\ref{app:sgr}). As this is the same as in the original SGR model, we will not discuss it further.

We next discuss the frequency dependence of the shear and odd viscosity spectra in Eqs.~(\ref{eta_shear},\ref{eta_odd}), focusing on the physically most interesting low-frequency regime $\omega,\Omega \ll 1/\tau_0$ and (in SGR units) $1<\mathcal{X}<2$, where the system is dense enough to be close to a glass transition. Using that then $\eta^*_0(\omega)\sim (i\omega)^{\mathcal{X}-2}$ ~\cite{sollich1997rheology,sollich1998rheological,cates2004tensorial,fielding2020elastoviscoplastic},
we find for frequencies $\omega \ll \Omega$ well below the active rotation frequency that $\eta_{\rm shear}^* \sim \eta_{\rm odd}^* \sim \Omega^{\mathcal{X}-2}$. The odd viscosity in a steady ($\omega\to 0$) shear flow thus {\em grows} as the active rotation frequency $\Omega$ decreases, a highly non-trivial result\footnote{The steady ($\omega\to 0$) {\em shear} viscosity from the inclusion stresses has the same scaling but is subleading compared to the matrix shear viscosity, which as in conventional SGR actually diverges for $1<\mathcal{X}<2$.}. {\red Intuitively, this behaviour comes from the power law spectrum of relaxation times in SGR, which causes the SGR viscosity $\eta_0^*(\omega)\sim (i\omega)^{\mathcal{X}-2}$ to have a diverging zero frequency limit. In the odd viscosity for the chiral SGR model, the active rotation with rate $\Omega$ can be viewed as providing an upper cutoff $\sim 1/\Omega$ on the relaxation times. As $\Omega$ decreases this cutoff grows, and so longer and longer relaxation timescales contribute to the odd viscosity, leading to the increase as $\Omega^{\mathcal{X}-2}$.}

\begin{figure}[htbp!]
    \centering
    \includegraphics[width=1.0\linewidth]{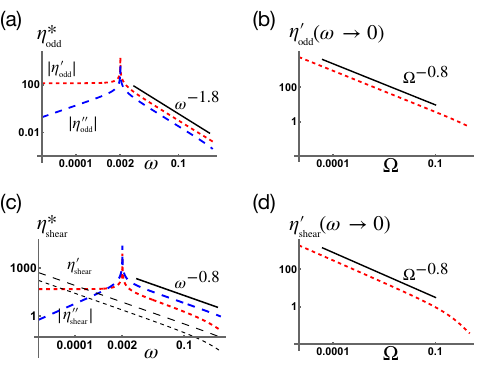}
\caption{\textbf{Odd and shear viscoelastic spectra of chiral SGR model}  -- (a) Real part $\eta_{\rm odd}'$ (red, short dashed) and imaginary part $\eta_{\rm odd}''$ (blue, long dashed) of odd viscosity against $\omega$, for $\Omega=0.001$ and $\mathcal{X}=1.2$. Black line: predicted power law scaling $\sim \omega^{\mathcal{X}-3}$ for $\omega\gg\Omega$. 
(b) The odd viscosity $\eta_{\rm odd}'(\omega\to 0)$ for steady shear, visible as the low frequency plateau in (a), scales as $\Omega^{\mathcal{X}-2}$ (black line).
(c,d) Analogs of (a,b) for the inclusion shear viscosity, which at $\omega\gg\Omega$ has the passive SGR scaling $\sim\omega^{\mathcal{X}-2}$ but a low-frequency plateau with the same $\Omega$-scaling as for the odd viscosity. Contributions from matrix stresses are shown by black dashed lines in (c).
Parameters: $\tau_0=1$, $k_1=1$, $k_2=0.05$.  {\red For our choice of parameters we find resonance peaks at the characteristic frequency $\omega=2\Omega=0.002$ in (a) and (c), which we have labeled explicitly on the x-axis.}}
    \label{fig:oddvis_sgr}
\end{figure}

In the opposite frequency regime $\omega \gg \Omega$ we find that the active rotation generates a frequency-dependent odd viscosity varying as $\eta^*_{\rm odd}(\omega)\sim \Omega \omega^{\mathcal{X}-3}$. {\red This vanishes as $\Omega\to 0$, ensuring that the odd viscosity at any fixed $\omega>0$ goes to zero in the passive, achiral limit of the model as expected.}
The $\Omega \omega^{\mathcal{X}-3}$ behaviour for $\omega\gg \Omega$ also matches the $\Omega^{\mathcal{X}-2}$ scaling for $\omega \ll \Omega$ when $\omega\approx \Omega$. Note that except for $\omega\to 0$, $\eta_{\rm odd}^*(\omega)$ has both real and imaginary parts, corresponding respectively to a non-vanishing odd viscous {\em and}  odd elastic response.
The plots in Fig.~\ref{fig:oddvis_sgr} (a) of the real and imaginary parts of the odd viscoelastic spectrum for $\mathcal{X} =1.2$ confirm the power law scaling $\Omega \omega^{\mathcal{X}-3}$ with frequency for $\omega\gg \Omega$. In the opposite frequency regime, Fig.~\ref{fig:oddvis_sgr}(b) demonstrates that the odd viscosity for steady flows, $\eta_{\rm odd}^*(\omega\to 0)$, which corresponds to the low-frequency plateau\footnote{
The plateau appears in the real part of the viscosity as expected; the imaginary part is smaller by a factor $\sim\omega/\Omega$.}
in Fig.~\ref{fig:oddvis_sgr}(a), has the expected $\Omega^{\mathcal{X}-2}$ scaling with $\Omega$.

Considering finally the crossover between the two regimes of $\omega$ well below and above $\Omega$, one sees from Fig.~\ref{fig:oddvis_sgr}(a) that, interestingly, this is not smooth but rather proceeds by a resonance-like peak at $\omega=2\Omega$. This can be rationalized intuitively as follows: an oscillatory shear flow creates extension and compression along axes at angles $\pm \pi/4$ to the flow, with extension and compression swapping roles every half period, i.e.\ after time $\pi/\omega$. If an inclusion has rotated by $\pi/2$ in this time, i.e.\ if $\pi/2 = \Omega(\pi/\omega)$, then any extension (respectively compression) accumulated in the previous half period is reinforced with every oscillation, effectively causing a resonance between external deformation and active rotation. 

The shear viscoelastic response of the inclusions as predicted by chiral SGR is plotted in Fig.~\ref{fig:oddvis_sgr}(c). As for the odd viscoelasticity it exhibits the predicted plateau of height $\sim \Omega^{\mathcal{X}-2}$ (Fig.~\ref{fig:oddvis_sgr}(d)) for $\omega\ll\Omega$ and a resonance peak at $\omega=2\Omega$. For $\omega\gg\Omega$, in contrast, the effects of activity become negligible in the shear viscoelasticity, consistent with Eq.~\eqref{eta_shear}, and one therefore observes the standard SGR frequency scaling $\sim \omega^{\mathcal{X}-2}$. 
Fig.~\ref{fig:oddvis_sgr}(c) additionally shows 
the matrix contribution to the shear viscoelasticity, which up to prefactors is identical to the result for passive SGR ($\Omega=0$). The fact that for $\omega\ll \Omega$ the shear viscoelasticity contributed by the inclusions lies below the contribution from the matrix (black dashed line) can be understood intuitively from the (in relative terms, fast) active rotation effectively averaging out compressional and extensional deformations. {\red Note that while in Fig.~\ref{fig:oddvis_sgr}
we have, for the sake of illustration, chosen specific values for the relevant elastic constants, we have checked that the qualitative behaviour remains the same for other values. Of course at least one of $k_1$ or $k_3$ has to be nonzero to ensure that the inclusions with their chiral properties contribute to the stress response.}

We comment briefly on the regime of high frequencies, although we expect in practice that other rheological contributions including e.g.\ solvent viscosity will be dominant in this regime. For $\omega\gg 1$, SGR predicts effectively Maxwell behaviour with a relaxation time of $\tau_0=1$, i.e.\ $G^*(\omega)\sim 1 + i\omega^{-1}$. Correspondingly, $\eta_0(\omega) \sim \omega^{-2}-i\omega^{-1}$ and one finds from Eqs.~(\ref{eta_shear},\ref{eta_odd}) by using $\Omega\ll \omega$
\begin{equation}
    \eta^*_{\rm shear}(\omega) \sim \omega^{-2} - i \omega^{-1} + O(\Omega^2), \qquad 
    \eta^*_{\rm odd}(\omega) \sim -\Omega \omega^{-2} - i \Omega\omega^{-3} + O(\Omega^3)
\end{equation}
Chiral activity thus again gives only small perturbations to the shear viscosity, while the odd viscosity is proportional to $\Omega$. For $\omega \sim 1$ these expressions connect smoothly to the ones for $\Omega \ll \omega \ll 1$,
\begin{eqnarray}
\eta^*_{\rm shear}(\omega) &=&\eta_0(\omega)+O(\Omega^2)
\sim 
\omega^{\mathcal{X}-2} - i \omega^{\mathcal{X}-2} 
\label{etashear_large_omega}
\\
\eta^*_{\rm odd}(\omega) &=&
2i\Omega \,\eta_0'(\omega) + O(\Omega^3)
\sim  -\Omega\,\omega^{\mathcal{X}-3} - i \Omega\, 
\omega^{\mathcal{X}-3}.
\label{etaodd_large_omega}
\end{eqnarray}

\section{Conclusions}

In this paper, we have developed a chiral SGR model as a mesoscopic description of actively rotating inclusions in a soft glassy matrix. We have analysed its predictions for the linear stress response to both steady and oscillatory shear flows. In both cases we find emergent odd terms in the response, giving an odd viscosity for steady flows and {\em odd viscoelasticity} in the oscillatory case. We find that, non-trivially, the odd viscosity can grow as the active rotation frequency $\Omega$ decreases towards zero. In oscillatory shear we generally find both odd viscous~\cite{ganeshan2017odd,banerjee2017odd,fruchart2023odd,souslov2019topological,Lier2023,duclut2024probe,banerjee2022hydrodynamic} and odd elastic~\cite{scheibner2020odd,fruchart2023odd,Starzewski2024,surowka2022odd} responses, 
which for frequencies $\omega\gg\Omega$
follow a power law dependence $\sim \Omega \omega^{\mathcal{X}-3}$. This regime of higher frequencies is separated from the low frequency range $\omega\ll\Omega$ by a resonance-like peak at $\omega=2\Omega$, where the active rotation reinforces extensional and compressional deformations in each period. 

In future work it would be interesting to explore the full nonlinear response of the chiral SGR model, for example to large amplitude oscillatory shear, which in recent years has been studied extensively in glassy systems~\cite{ewoldt2008new,Parley2022,radhakrishnan2016shear,radhakrishnan2018shear}. We expect nonlinear effects to be particularly important around the resonance peaks we have identified, or generally in settings where large residual deformations remain after yield events. {\red A study of nonlinear effects would of course also require one to revisit the simple quadratic form we have assumed for the elastic energy of mesoscopic elements, including careful consideration of the boundary conditions between inclusions and matrix. Likewise the accuracy of the pre-averaging approximation in such a nonlinear regime would have to be checked explicitly by comparison with the full model.}
It will also be interesting to compare the model predictions to realizations in particle systems, e.g.\ in dense mixtures of passive particles and active spinners. {\red 
Given that odd viscosity is known to show an even richer behavior in three rather than two dimensions~\cite{hosaka2023lorentz,hosaka2024chirotactic,khain2022stokes}, a further direction of interest would be an extension of the model from two to three dimensions. Here the active rotation rate would be specified by a (pseudo-)vector $\bm{\Omega}$. Including a stochastic component in the active rotation as is done in chiral active Brownian particle models~\cite{debets2023glassy,han2021fluctuating,zhang2026unusual,caprini2026modeling,sevilla2016diffusion} would also be interesting; in the three-dimensional version of this model this could then also allow stochastic changes in the direction of $\bm{\Omega}$.} 

{\red Our model is obviously simplified but can provide a useful basis from which to explore odd viscoelastic effects arising from chirality in experimental systems. Indeed, one inspiration for our model is given by observations on biological systems where the cell nucleus rotates actively~\cite{kumar2014actomyosin}. More recently, odd elasticity has also been measured in chirally rotating starfish embryos~\cite{tan2022odd}; at higher densities these embryos can be expected to behave as odd viscoelastic materials. Related work has extended odd elasticity to muscle fibers~\cite{shankar2024active}. Active viscoelastic behavior has also been reported in epithelial tissues~\cite{ioratim2023mechanochemical}, which could likewise be generalized to chiral systems to reveal odd mechanical effects.}
Overall, the combination of glassy effects as captured by SGR and chirality in active matter provides a rich ground for the theoretical study of {\em odd viscoelastic} responses and a new approach to the modelling of materials with active torques~\cite{banerjee2021active,surowka2022odd,duclut2024probe,soni2019odd,hargus2024odd}. Given the prevalence of 
chirality both at biomolecular level~\cite{compton2002chirality} 
and on larger scales in the motion of microorganisms in active glasses~\cite{lama2024emergence}, we believe that studies in this direction will provide a deeper understanding of experimentally realizable chiral active systems~\cite{huang2020dynamical,furthauer2012active,furthauer2013active} in the near future, for which the chiral SGR model can provide paradigmatic theoretical insights.

\ack{DB acknowledges Ananyo Maitra, Andrej Vilfan and Rituparno Mandal for constructive discussions. PS acknowledges funding by the Deutsche Forschungsgemeinschaft (DFG, German Research Foundation), Project-ID 449750155, RTG 2756, Project A5.} 

\bibliography{bibliography.bib}

\appendix

\section{Time evolution of $G(E,t)$}
\label{app:G}

{\red In this appendix we give details of how to integrate Eq.~\eqref{full} over the deformation variables to obtain Eq.~\eqref{G_dot} for the time evolution of $G(E,t)$. The first term in Eq.~\eqref{full} gives upon integration
\begin{align}
    \left( \int d {\bm F}^I d {\bm F}^M P_1 \right) \frac{\partial}{\partial t} G(E,t) = \frac{\partial}{\partial t} G(E,t)
\end{align}
because $P_1\equiv P_1(\bm{F}^I,\bm{F}^M|E,t)$ is a normalized probability distribution. For the same reason the second term vanishes:
\begin{align}
    G(E,t) \frac{\partial}{\partial t} \left( \int d {\bm F}^I d {\bm F}^M P_1 \right) = 0\ .
\end{align}
From the third term one obtains
\begin{align}
    \int d {\bm F}^I d {\bm F}^M G(E,t)  \frac{\partial}{\partial F_{ij}^I} \left[ (\partial_k v_i) F_{kj}^I P_1 \right] = G(E,t) (\partial_k v_i) \int d {\bm F}^I d {\bm F}^M \frac{\partial}{\partial F_{ij}^I} \left( F_{kj}^I P_1 \right)
\end{align}
The integral that remains here gives only boundary terms at $F_{ij}^I=\pm \infty$, where $P^1$ as an integrable distribution must vanish, so the result is zero. The fourth term similarly only produces boundary terms after integration:
\begin{align}
    \int d {\bm F}^I d {\bm F}^M G(E,t) \frac{\partial}{\partial F_{ij}^I} \left(\Omega \epsilon_{ik} F_{kj}^I  P_1 \right) = G(E,t) \Omega \epsilon_{ik} \int d {\bm F}^I d {\bm F}^M \frac{\partial}{\partial F_{ij}^I} \left(F_{kj}^I  P_1 \right)\ .
\end{align} 
The fifth integral looks exactly like the third one but with ${\bm F}^I$ replaced by ${\bm F}^M$, so vanishes for the same reason.}

{\red In the first term on the right hand side of Eq.~\eqref{full}, the integral of $f P_1$ over ${\bm F}^I$ and ${\bm F}^M$ just defines the average written in the first term on the r.h.s.\ of Eq.~\eqref{G_dot}. Finally, the second term on the r.h.s.\ of Eq.~\eqref{full} yields after integration
\begin{align}
    \int d {\bm F}^I d {\bm F}^M \Gamma(t) \rho(E) 
    \rho^I\!(\bm{F}^I)\rho^M\!(\bm{F}^M) = \Gamma(t) \rho(E) \left( \int d {\bm F}^I \rho^I({\bm F}^I) \right) \left( \int d {\bm F}^M \rho^M({\bm F}^M) \right).
\end{align}
and the last two integrals are unity because the probability distributions $\rho^I$ and $\rho^M$ are normalised.}

\section{Time evolution of $\sigma^M_{ij}(E,t)$
}
\label{app:premultiply}

{\red As explained in the main text, in order to obtain Eq.~\eqref{sigmaM_dot} for the time evolution of $\sigma^M_{ij}(E,t)$ one has to multiply Eq.~\eqref{full} by $l_{ij}^M$ and integrate over $\bm{F}^I$ and $\bm{F}^M$. We give details here of this calculation for the fifth term on the l.h.s.\ of Eq.~\eqref{full}.
Multiplying this term by 
$l_{ij}^M = (F_{ik}^M F_{jk}^M-\delta_{ij})/2$ and integrating by parts gives
\begin{align}
& \frac{G(E,t)}{2} \int d\bm{F}^I d\bm{F}^M (F_{ik}^M F_{jk}^M -\delta_{ij})\frac{\partial }{\partial F_{lm}^M} \left((\partial_n v_l) F_{nm}^M P_1 \right)  = \nonumber \\
&= - \frac{G(E,t)}{2} \left\langle  (\partial_n v_l) F_{nm}^M \left( \frac{\partial }{\partial F_{lm}^M} F_{ik}^M F_{jk}^M\right) \right\rangle_{E,t} \\
    & = -\frac{G(E,t)}{2} \left\langle \left[ (\partial_n v_i) F_{nk}^M F_{jk}^M + (\partial_n v_j) F_{nk}^M F_{ik}^M \right] \right\rangle_{E,t} \nonumber \\
    & = - \frac{G(E,t)}{2} \left\langle \left[ (\partial_n v_i) ( 2l_{nj}^M + \delta_{nj}) + (\partial_n v_j) ( 2l_{ni}^M + \delta_{ni}) \right]\right\rangle_{E,t} \nonumber \\
    & = - \left[(\partial_k v_i) \sigma_{kj}^M(E,t)/k_2 +  (\partial_k v_j) \sigma_{ki}^M(E,t)/k_2 + G(E,t) v_{ij}  \right]\ . \nonumber
\end{align}
which is the term in square brackets on the l.h.s.\ of Eq.~\eqref{sigmaM_dot}. As in the main text, $v_{ij}=(\partial_i v_j + \partial_j v_i)/2$ in the above expression is the symmetrized velocity-gradient tensor.}

\section{Matrix contribution to viscoelastic spectra}
\label{app:sgr}

{\red Here we briefly discuss the effect of the matrix on the viscoelastic response. 
Writing out the factor $\tau_\omega$ in Eq.~\eqref{eq:vis_response} gives
\begin{align}
    \sigma_{ij}^M(E) = \frac{k_2 G(E) \tau(E)}{1 + i \omega \tau(E)} v_{ij}
\end{align}
where $\sigma^M_{ij}$, $G$ and $\tau$ are functions of the yield energy $E$ as indicated.
Integrating over $E$ as in the main text leads to a total matrix stress of 
\begin{align}
    \sigma_{ij}^M =  
    k_2 \int_0^{\infty} d E \frac{G(E) \tau(E)}{1 + i \omega \tau (E)} \, v_{ij} = \frac{k_2}{k_1}\eta_0^*(\omega) \,v_{ij}
\end{align}
where the last equality follows by comparison with Eq.~\eqref{eta0_int}. The proportionality of $\sigma_{ij}^M$ to $v_{ij}$ shows that no odd viscoelastic terms appear, while the contribution to the shear viscosity is proportional to $\eta_0^*(\omega)$ as stated in the main text.
}

\section{Comparison with symmetry-based models of odd viscoelasticity}
\label{app:oddspectra}

\newcommand{\etae}{\eta^{\rm e}}
\newcommand{\etao}{\eta^{\rm o}}
\newcommand{\ko}{\kappa^{\rm o}}

{\red We discuss here the relation between the chiral SGR model and the odd Maxwell model discussed in~\cite{banerjee2021active}. The latter does not have glassy features, but can be compared to the relation between stress and strain in the chiral SGR model at {\em fixed} yield energy $E$, specifically Eq.~\eqref{sigma1_SM_lin}. The odd Maxwell model from~\cite{banerjee2021active} in its linearized form relates the strain rate (i.e.\ symmetrized velocity gradient) and stress tensors via
\begin{equation}
\left(
\!\!\begin{array}{c}    
v_{xx}\\v_{yy}\\v_{xy}
\end{array}\!\!
\right)
=
\left(\!\!
\begin{array}{ccc} 
2\etae + \zeta\!\! & \zeta & 2\etao \\
\,\,\zeta\!\! & \!\!2\etae + \zeta\, & \!\!\!\!-2\etao \\
\,-\etao\! & \,\etao\! & 2\etae
\end{array}
\!\!\right)^{-1}
\left(\!\!
\begin{array}{c}
\sigma_{xx}\\\sigma_{yy}\\\sigma_{xy}
\end{array}\!\!
\right)
+ 
\left(\!\!
\begin{array}{ccc}    
2\mu + \lambda\!\! & \lambda & 2\ko \\
\,\,\lambda\!\! & \!\!2\mu + \lambda\, & \!\!\!\!-2\ko \\
\,-\ko\! & \,\ko\! & \!2\mu
\end{array}
\!\!\right)^{-1}
\frac{\partial}{\partial t}
\left(
\!\!
\begin{array}{c}    
\sigma_{xx}\\\sigma_{yy}\\\sigma_{xy}
\end{array}
\!\!
\right)
\label{odd_Maxwell}\end{equation}
The matrices appearing on the r.h.s.\ are a generic viscosity and elasticity tensor, respectively, translated from Ref.~\cite{banerjee2021active} into our $3\times 3$ matrix notation.
Specifically, $\mu$ and $\lambda$ are the conventional Lam\'e elastic constants for shear and compression, while $\ko$ is an additional elastic constant that represents odd elastic effects. Similarly in the viscosity tensor, $\etae$ and $\zeta$ are conventional shear and bulk viscosity parameters, while $\etao$ allows for an additional odd viscous response. Note that $\etae$ is denoted $\eta$ in~\cite{banerjee2021active}; we use the ``e'' (even) superscript here to distinguish it from other occurrences of $\eta$.

The form of Eq.~\eqref{odd_Maxwell} was derived in~\cite{banerjee2021active} based only on the assumptions of isotropy and symmetry of stress and strain rate tensors. The chiral SGR model includes the same assumptions so Eq.~\eqref{sigma1_SM_lin} must be consistent with the general form of Eq.~\eqref{odd_Maxwell}. This is indeed so: one checks by a direct calculation that Eq.~\eqref{odd_Maxwell} turns into Eq.~\eqref{sigma1_SM_lin} if the parameters are chosen as
\begin{eqnarray}
2\mu = k_1 G, \quad \lambda=\ko=0, \quad 
\etae = \frac{k_1 G \tau}{2 (1 + 4 \Omega^2 \tau^2)}, \quad
\etao=2 \Omega \tau \,\etae,
\quad 
\zeta=
4\Omega^2\tau^2\etae
\end{eqnarray}
This correspondence confirms that in chiral SGR the local {\em elasticity} is assumed to be of conventional even type ($\ko=0$); the odd effects in the response at fixed yield energy $E$ come from the {\em viscous} term $\etao$, which as expected vanishes in the achiral limit $\Omega\to 0$.
\par
Conversely, one can ask what viscoelastic spectra are obtained from the odd Maxwell model in Eq.~\eqref{odd_Maxwell} for generic parameter choices. By a calculation that is fully analogous to the one in Sec.~\ref{sec:oscillatory} and therefore not detailed here, one finds that the frequency-dependent shear and odd viscosities are given by expressions of the same form as the fixed-$E$ results for chiral SGR, Eqs.~\eqref{shear_via_plus_minus} and~\eqref{odd_via_plus_minus}. The corresponding $\eta_\pm$ are 
\begin{equation}
\eta_\pm = \frac{(\mu+i\ko) \tilde\tau}{1 + i \omega \tilde\tau  \pm 2 i \tilde\Omega \tilde\tau}
    \label{etapm_Maxwell}
\end{equation}
with an effective relaxation time $\tilde\tau$ and rotation frequency $\tilde\Omega$ given by
\begin{eqnarray}
\tilde\tau = \frac{(\etae)^2 + (\etao)^2}{\etae\mu + \etao \ko}, \qquad
\tilde\Omega = \frac{\etae \ko - \etao \mu}{2 [(\etae)^2 + (\etao)^2]}
\end{eqnarray}
Comparing with Eq.~\eqref{eta0pm_def} shows that these expressions for $\eta_\pm$ are identical to those in chiral SGR at fixed $E$, except for the replacements of $k_1 G$ by $2(\mu+i\ko)$, of $\tau$ by $\tilde\tau$ and of $\Omega$ by $\tilde\Omega$. Using the intuition gained from the construction of the chiral SGR we therefore deduce that, in the odd Maxwell model of~\cite{banerjee2021active}, the viscous stress relaxation can be interpreted as a combination of a conventional Maxwell relaxation on timescale $\tilde\tau$ and a stress tensor rotation at angular frequency $\tilde\Omega$. For vanishing odd elasticity ($\ko=0$), where the numerator of the r.h.s.\ of Eq.~\eqref{etapm_Maxwell} is real, the frequency-dependent shear and odd viscosities are then in fact quantitatively the same as in fixed-$E$ chiral SGR with appropriately matched parameters; for $\ko\neq 0$ the additional imaginary term in the numerator effectively mixes the real and imaginary parts of the viscoelastic spectra.

We finally illustrate the qualitative frequency dependence of the shear and odd viscosities for the odd Maxwell model as given by Eqs.~(\ref{shear_via_plus_minus},\ref{odd_via_plus_minus},\ref{etapm_Maxwell}), for the case without odd elasticity ($\ko=0$). As a function of scaled frequency $\omega/\tilde\Omega$, the shape of these functions is then determined by the dimensionless parameter $\tilde\Omega\tilde\tau$. When this is $\ll 1$, the shear viscosity is largely unaffected by odd effects and follows conventional Maxwell behaviour; the odd viscosity is then small in comparison (see Fig.~\ref{fig:oddMax}(b,d)). For $\tilde\Omega\tilde\tau \gtrsim 1$, on the other hand, a near-resonance becomes visible, at a frequency in $\omega=2\tilde\Omega$ matching the one in fixed-$E$ chiral SGR (see Fig.~\ref{fig:oddMax}(a,c)). Qualitative differences to the full chiral SGR prediction of course remain nonetheless, because the odd Maxwell model~\cite{banerjee2021active} does not have glassy features. In chiral SGR these arise from the averaging over the distribution of yield energies $E$ or equivalently over a power law distribution of relaxation times $\tau$. This leads in particular to the resonance at $\omega=2\tilde\Omega$ turning into a divergence (when $1<\mathcal{X}<2$), as well as fractional power law dependences on $\omega$ as discussed in Sec.~\ref{sec:oscillatory}. These differences in behaviour between the full chiral SGR model and the odd Maxwell model can be appreciated from a comparison of Fig.~\ref{fig:oddvis_sgr} (a,c) with Fig.~\ref{fig:oddMax} (a,c).

\begin{figure}
    \centering
    \includegraphics[width=1.0\linewidth]{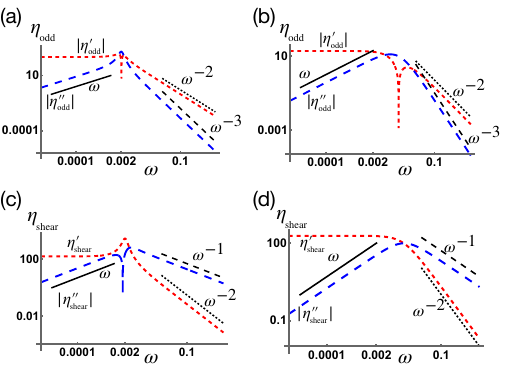}
    \caption{\textbf{Odd and shear viscoelastic spectra of odd Maxwell model}  -- (a) Real part $\eta_{\rm odd}'$ (red, short dashed) and imaginary part $\eta_{\rm odd}''$ (blue, long dashed) of odd viscosity against $\omega$, for $\mu=1$, $\ko=0$, $\tilde\Omega=0.001$ and $\tilde\tau=3000$, i.e.\ $\tilde\Omega\tilde\tau=3$. (b) As (a) but for $\tilde\tau=100$ ($\tilde\Omega\tilde\tau=0.1$).  (c,d) Analogues of (a,b) for the shear viscosity. 
    Black solid and dashed/dotted lines indicate the scalings in the odd Maxwell model for $\omega\ll \tilde\Omega$ and $\omega\gg \tilde\Omega$, respectively; the latter differ from the glassy scalings in chiral SGR, cf.\ Fig.~\ref{fig:oddvis_sgr}(a,c). For ease of comparison, the values of $\tilde\Omega$ here and $\Omega$ in Fig.~\ref{fig:oddvis_sgr} have been chosen to be the same (0.001), to allow a direct comparison. 
    At $\omega=2\tilde\Omega$ the odd Maxwell shows resonance-like peaks if $\tilde\Omega\tilde\tau\gtrsim 1$ as in (a,c). In chiral SGR (Fig.~\ref{fig:oddvis_sgr}(a,c)) these peaks become divergences arising from the power law relaxation time spectrum. While only absolute values are plotted, the signs of all response functions match those of chiral SGR in the different frequency regimes. In the odd Maxwell model the sign changes occur without divergences, giving the cusp-like structures in the absolute value plots.
    }
    \label{fig:oddMax}
\end{figure}
}

\section{Energy Balance}
\label{app:energy}

We comment briefly on the energy balance in the chiral SGR model defined in the main text. To find this we can multiply Eq.~\eqref{eq:sgr} by $\Eel$ and integrate over $\bm{F}^I$, $\bm{F}^M$ and $E$. After appropriately integrating by parts this gives 
\begin{align}
(d/dt)\langle \Eel\rangle &= (\partial_k v_i)
\left\langle F^M_{kj} \frac{\partial \Eel}{\partial F^M_{ij}}
\right\rangle 
+
(-\Omega\epsilon_{ik}+ \partial_k v_i)\left\langle F^I_{kj} \frac{\partial \Eel}{\partial F^I_{ij}}
\right\rangle  \nonumber \\
&-\Gamma_0 \langle 
\Eel e^{-(E-\Eel)/\mathcal{X}}\rangle + \Gamma(t)\langle \Eel\rangle_\rho
\label{balance}
\end{align}
where the last term represents the average elastic energy stored locally after a yield, 
which can be nonzero due to residual deformations. From the definition of the Green-Lagrange strain tensor (Eq.~\eqref{Green-Lagrange}) we get the Jacobian in going from deformation to strain:
\begin{align}
    \frac{\partial l_{ij}^M}{\partial F_{mn}^M} & = \frac{1}{2} \left( F_{ik}^M \frac{\partial F_{jk}^M}{\partial F_{ij}^M} + \frac{\partial F_{ik}^M}{\partial F_{ij}^M} F_{jk}^M \right) \nonumber \\ 
    &= \frac{1}{2} \left( \delta_{im} \delta_{ln} F_{jl}^M + \delta_{jm} \delta_{ln} F_{il}^M \right)
\end{align}
The derivatives w.r.t.\ the deformation tensors can be expressed as ones w.r.t.\ the strains as follows:
\begin{equation}
    F^M_{kj} \frac{\partial \Eel}{\partial F^M_{ij}} = 
(2l^M_{km}+\delta_{km})
\frac{\partial \Eel}{\partial l^M_{im}}
\end{equation}
and similarly for $F^I_{ij}$. This gives for the first term on the r.h.s.\ of~\eqref{balance}
\begin{equation}
(\partial_k v_i) \left\langle 
2l^M_{km}
\frac{\partial \Eel}{\partial l^M_{im}} + 
\frac{\partial \Eel}{\partial l^M_{ik}}
\right\rangle 
\end{equation}
For small strains the first part can be neglected. Proceeding similarly for the second term in~\eqref{balance} gives
\begin{align}
    (d/dt)\langle \Eel\rangle &= (\partial_k v_i)
\left\langle \frac{\partial \Eel}{\partial l^M_{ik}}+ \frac{\partial \Eel}{\partial l^I_{ik}}\right \rangle 
-\Omega\epsilon_{ik}\left\langle 
\frac{\partial \Eel}{\partial l^I_{ik}}\right\rangle \nonumber \\
& -\Gamma_0 \langle 
\Eel e^{-(E-\Eel)/\mathcal{X}}\rangle + \Gamma(t)\langle \Eel\rangle_\rho
\label{balance1}
\end{align}
One recognizes in the first term $(\partial_k v_i)\sigma_{ik} = v_{ik}\sigma_{ik}$, which is the rate of mechanical work expressed in terms of total stress and (symmetrized) velocity gradient. The last two terms give the dissipation rate from yield events. Generalizing slightly from the original SGR model~\cite{sollich1998rheological}, one sees that the model effectively assumes that the dissipated energy in a yield is the difference between elastic energy before the yield (penultimate term) and after the yield (from residual deformations, last term). The second term from the active rotation vanishes because of the antisymmetry of $\epsilon_{ik}$, so that the overall energy balance relation has the same structure as in SGR, i.e.\ the rate of change of elastic energy (l.h.s.) is the difference of mechanical work (first term on r.h.s.) and dissipated energy (last two terms on r.h.s.).
\end{document}